\documentstyle[sprocl,epsfig]{article}

\input{psfig}

\def\be{\begin{equation}}
\def\ee{\end{equation}}
\def\be{\begin{equation}}
\def\ee{\end{equation}}
\def\bq{\begin{eqnarray}}
\def\eq{\end{eqnarray}}

\begin{document}

\begin{flushright}
WUE-ITP-98-032\\
\end{flushright}
\bigskip

\title{EXCLUSIVE NONLEPTONIC DECAYS OF HEAVY MESONS IN QCD 
\footnote
{{\it talk presented by A. Khodjamirian at the 3rd Workshop 
``Continuous Advances in QCD'', Minneapolis, April 1998}}
}

\author{A. KHODJAMIRIAN \footnote{{\small on 
leave from Yerevan Physics Institute, 
375036 Yerevan, Armenia }}, R. R\"UCKL}

\address{ Institut f\"ur Theoretische Physik,
Universit\"at W\"urzburg,\\ D-97074 W\"urzburg,
Germany\\E-mail: ak@physik.uni-wuerzburg.de, 
rueckl@physik.uni-wuerzburg.de }

\maketitle\abstracts{ 
We use operator product expansion and QCD light-cone sum rules
to estimate the amplitude of the decay 
$B \to J/\psi K $  taking into account leading 
nonfactorizable contributions. The result is very similar
to the estimate obtained earlier from a four-point QCD sum rule.
We discuss applications of this method to other 
nonleptonic $B$  decay modes.}

\section{Introduction}
Exclusive nonleptonic decays of $B$ and $D$ mesons are complicated 
processes influenced by strong interactions 
at small and large distances. Theoretically, the decay  
amplitudes are determined from an effective weak Hamiltonian in terms
of short-distance Wilson coefficients and matrix elements of local 
four-quark operators. While the 
short-distance effects can be calculated in QCD 
perturbation theory using renormalization group methods,  it is 
extremely difficult to obtain reliable and accurate estimates for 
the hadronic matrix elements.
Here we report on an estimate of 
the matrix element for  the 
decay $B\to J/\psi K$ using QCD sum rule techniques. 

The decay amplitude for $B \rightarrow J/\psi K $ is determined 
by the matrix element 
$$
\langle K J/\psi \mid H_W  \mid B\rangle
 = \frac{G_F}{\sqrt{2}}V_{cb}V_{cs}^*
\Bigg [\left(c_2(\mu)+\frac{c_1(\mu)}3\right)
\langle K J/\psi \mid O_2(\mu) \mid B\rangle 
$$
\be
+2c_1(\mu)
\langle K J/\psi \mid \widetilde{O}_2(\mu) \mid B\rangle \Bigg]~,
\label{ampl}
\ee
where $G_F$ is the Fermi constant, $V_{cb}$ and $V_{cs}$ are the 
relevant CKM matrix elements, and   
$c_1(\mu)$ and $c_2(\mu)$ are the Wilson coefficients 
of the four-quark operators  
\be
O_1=(\bar{s}\Gamma^\rho c)(\bar{c}\Gamma_\rho b),~~
O_2=(\bar{c}\Gamma^\rho c)(\bar{s}\Gamma_\rho b),
\label{01}
\ee
respectively, with $\Gamma_\rho=\gamma_\rho(1-\gamma_5)$.
The operator 
\be
\widetilde{O}_2=(\bar{c}\Gamma^\rho \frac{\lambda^a}{2}c)
(\bar{s}\Gamma_\rho \frac{\lambda^a}{2}b)
\label{Otilde}
\ee
with $tr(\lambda^a\lambda^b)=2\delta^{ab}$ 
originates from the Fierz rearrangement of $O_1$.

In the usual factorization approximation, 
the matrix 
element of $O_2$ is split into the  product
\be
\langle
K J/\psi \mid O_2(\mu)\mid B\rangle
 =  
\langle J/\psi \mid \bar{c}\Gamma^\rho c  \mid 0 \rangle
\langle K  \mid \bar{s}\Gamma_\rho b \mid B\rangle ~,
\label{fact}
\ee
involving simpler matrix elements of quark currents:
\be
\langle 0 \mid \bar{c}\Gamma^\rho c \mid J/\psi(p)\rangle= 
f_\psi m_{\psi}\epsilon^\rho,
\label{constpsi}
\ee
\be
\langle K(q) \mid \bar{s}\Gamma_\rho b \mid B(p+q)\rangle
 = 2f^+(p^2)q_\rho + \left(f^+(p^2)+f^-(p^2)\right)p_\rho ~.    
\label{formf}
\ee
In the above,
$f_\psi$ is the $J/\psi$ decay constant, 
$\epsilon^\rho$ is the polarization vector,
and $f^{\pm}(p^2)$ are form factors.   
The momentum assignment is also indicated.
The matrix element of 
$\widetilde{O}_2$ vanishes in this approximation 
because of colour conservation.

The short-distance coefficients $c_{1,2}(\mu)$ 
and the matrix elements entering (\ref{ampl}) are scale-dependent, whereas 
the decay constants and form factors determining the r.h.s. of (\ref{fact}) 
are physical scale-independent quantities. Therefore,   
factorization can at best be an approximation valid 
at a particular scale.   
As shown in ref. \cite{KR},
in the $B\to J/\psi K$  channel factorization does not work at 
$\mu = {\cal O}(m_b)$. 
The main nonfactorizable contribution comes from the 
matrix element of $\widetilde{O}_2$ which is 
conveniently parametrized \cite{KR} in the form  
\be
\langle
K J/\psi \mid\widetilde{O}_2(\mu) \mid B \rangle
= 2f_\psi m_{\psi}\tilde{f}_{B\psi K}(\mu)(\epsilon \cdot q)  ~.
\label{def}
\ee
Here, the dimensionless quantity $\tilde{f}_{B\psi K}(\mu)$
is the analog of $f^+(m_\psi^2)$ in (\ref{fact}).
Note that the nonfactorizable effects in the matrix element of 
$O_1$ are subdominant and therefore neglected here.
Substitution of  (\ref{fact}) and (\ref{def}) 
in (\ref{ampl}) yields 
\be
\langle K J/\psi \mid H_W  \mid B\rangle = 
\sqrt{2}G_F V_{cb}V_{cs}^*a_2^{B\psi K}f_\psi
f^+m_{\psi}( \epsilon \cdot q)~
\label{decay}
\ee
with
\be
a_2^{B\psi K}=c_2(\mu)+\frac{c_1(\mu)}3 + 
2c_1(\mu)\frac{\tilde{f}_{B\psi K}(\mu)}{f^+(m_\psi^2)}~.
\label{a2}
\ee

An estimate
of the matrix element (\ref{def}) using four-point QCD sum rules 
is described in ref. \cite{KR}.
In this approach, originally suggested in ref. \cite{BS87} for $D$ decays,
gluonic interactions which break factorization 
are associated 
with the nonperturbative nature of the QCD vacuum as 
described by quark and gluon condensates. 
The nonfactorizable matrix element (\ref{def}) 
turns out  to be  small in comparison to the factorized matrix element 
(\ref{fact}). 
Numerically, one finds 
\be
\tilde{f}_{B\psi K}(\mu) = -(0.045 ~\mbox{to}~ 0.075)
\label{number4pt}
\ee
and $f^+(m_\psi^2)=0.55$. Note that the relevant scale $\mu$ in 
(\ref{number4pt}) is of 
order of the Borel mass $M \simeq \sqrt{m_B^2-m_b^2} \simeq 2.4$ GeV. 
Although the ratio $\tilde{f}_{B\psi K}(M)/f^+(m_\psi^2)$ $\simeq 0.1$  
is small,  
it has a strong impact in (\ref{a2}) because of the enhancement by the large short-distance 
coefficient $c_1(M) \simeq 1.1$ and because of a partial cancellation 
in  $c_2(M)+c_1(M)/3 \simeq 0.09$. The resulting prediction for 
$a_2$ is comparable in size
but opposite in sign to the outcome of phenomenological fits \cite{BSW,NS} 
assuming a universal coefficient $a_2$ (and $a_1$). It should be 
stressed that in our approach the nonfactorizable effects are 
expected to be nonuniversal. Furthermore, the nonfactorizable 
term in (\ref{a2}) tends to cancel 
the nonleading in $1/N_c$ factorizable term $c_1/3$. This resembles 
the $1/N_c$-rule for $D$ decays \cite{BGR}.

An alternative approach was suggested 
in ref. \cite{BS93} using operator
product expansion (OPE) and absorbing the nonfactorizable gluon interactions 
in hadronic matrix elements
of new effective operators.  In ref. \cite{BS93} this method was 
applied to the decay mode $\bar{B}^0\to D^+ \pi^-$. 
In the heavy quark limit, the new operator reduces
in this case to the HQET chromomagnetic operator. The matrix element
of the latter between $B$ and $D$ states can be reliably 
estimated. A similar approach can also be  
applied \cite{Halperin} to decay modes where the nonfactorizable amplitudes 
involve hadronic matrix elements between a $B$ and a light meson. 
Here we describe such an estimate for $B\rightarrow J/\psi K$.

\section{The correlation function and OPE}

As suggested in ref. \cite{BS93}, 
we replace the $J/\psi$ state in (\ref{def}) by the generating 
current $\bar{c}\gamma_\mu c$ and consider the following 
correlation function:
$$
A_{\mu} (p,q)=
i\int d^4x ~e^{ipx}
\langle K(q) \mid T\{\bar{c}(x)\gamma_\mu c(x),\widetilde{O}_2(0)\}\mid 
B(p+q)\rangle 
$$
\be
= \left( -p^2q_\mu +(p\cdot q)p_\mu \right)
A(p^2)~, 
\label{corr}
\ee
where $(p+q)^2=m_B^2$ and $q^2 = m_K^2$ is put to zero. 
Because of current conservation, there is only 
one invariant amplitude $A$.   
For the momentum $p$ of the $\bar{c}\gamma_\mu c$ current 
we require 
\be
p^2 \ll 4m_c^2.
\label{region}
\ee 

Inserting a complete set of states 
with $J/\psi$ quantum numbers between the operators in (\ref{corr}) 
one gets, schematically, 
\be
A_\mu(p,q)= 
\frac{\langle 0|  
\bar{c}\gamma_\mu c| J/\psi 
\rangle \langle K J/\psi |\widetilde{O}_2|
B\rangle}{m_\psi^2 -p^2}
+\sum_{h= \psi',...}
\frac{\langle 0|  
\bar{c}\gamma_\mu c| h 
\rangle \langle K h |\widetilde{O}_2|
B\rangle}{m_h^2 -p^2}~.
\label{disp}
\ee
Substitution of 
(\ref{constpsi}) and (\ref{def})
in (\ref{disp}) then leads to the following dispersion relation for 
the invariant amplitude $A$: 
\be
A(p^2) =\frac{2f_\psi^2\tilde{f}_{B\psi K}}{m_\psi^2-p^2}
+\frac{2f_{\psi'}^2\tilde{f}_{B\psi' K}}{m_{\psi'}^2-p^2}
+\int\limits_{4m_D^2}^{\infty}
\frac{\rho^h (s)ds}{s-p^2}~.
\label{disppsi}
\ee
The $J/\psi$ ground-state contribution in this relation
contains the desired matrix element (\ref{def}).
The second term in the r.h.s. represents the contribution of the 
$\psi'$ resonance, whereas the dispersion integral 
takes into account all excited charmonium and non-resonant states 
in $J/\psi$ channel located above the open charm threshold.
Possible subtraction terms are not shown for brevity. 
They will be eliminated later.
In the region (\ref{region}), it is possible 
to expand the product of operators entering (\ref{corr})
in a series of local operators. A similar expansion 
has been applied in ref. \cite{KRSW} in order to estimate the long-distance 
effect in $B\to K^* \gamma$ due to interaction of virtual charmed quarks
with soft gluons. As a first step, we contract the $c$-quark fields 
in the product of  the currents 
$\bar{c}\gamma_\mu c$ and 
$\bar{c}\Gamma^\rho \frac{\lambda^a}{2}c$ and use 
$$
\langle 0|T\{c(x)\bar{c}(0)\}|0\rangle = 
i\int \frac{d^4k}{(2\pi)^4}e^{-ik(x-y)}\Bigg\{
\frac{\not\!k+m_c}{k^2-m_c^2}+
$$
$$
+\frac12g_sG^a_{\tau\lambda}(0)\frac{\lambda^a}2
\Big[ \frac {\epsilon^{\tau\lambda\delta\beta}\gamma_\delta\gamma_5k_\beta-
m_c\sigma^{\tau\lambda}}{(k^2-m_c^2)^2} 
$$
\be
+ y^\tau\frac{(k^2-m_c^2)\gamma^\lambda -2k^\lambda(\not\!k+m_c)}{(k^2-m_c^2)^2}\Big] \Bigg\}~.
\label{prop}
\ee
Here the first term is the free $c$-quark propagator and the second 
term incorporates the one-gluon emission
in the fixed-point ($x=0$) gauge $A_\mu^a(x) =1/2x^\tau G^a_{\tau\mu}(0)$.
The free-quark loop vanishes because of $Tr\lambda^a=0$.
The lowest order nonvanishing contribution involves 
a single gluon-field tensor. 
The corresponding Wilson coefficient can be calculated from 
the diagrams shown in Fig. 1. 
The result is :
\be
i\int d^4x e^{ipx}T\{ \bar{c}(x)\gamma_\mu c(x),\bar{c}\Gamma_\nu
\frac{\lambda^a}2 c(0)\} = t_{\mu\nu\alpha\beta}g_s\widetilde{G}^{a\alpha\beta}
I_c(p^2) + ...
\label{expan}
\ee   
with
\be
t_{\mu\nu\alpha\beta}= p_\mu p_\alpha g_{\nu\beta}+
p_\nu p_\alpha g_{\mu\beta}-p^2g_{\mu\alpha}g_{\nu\beta} ~,
\ee

\be
\widetilde{G}^a_{\alpha\beta}= \frac12
\epsilon _{\alpha\beta \sigma\tau}G^{a\sigma\tau}~, 
\ee

\be
I_c(p^2) = \frac1{4\pi^2}\int\limits^1_0 dx \frac{x^2(1-x)}{m_c^2-p^2x(1-x)}~.
\ee

\begin{figure}[hb]
\mbox{
\epsfig{file=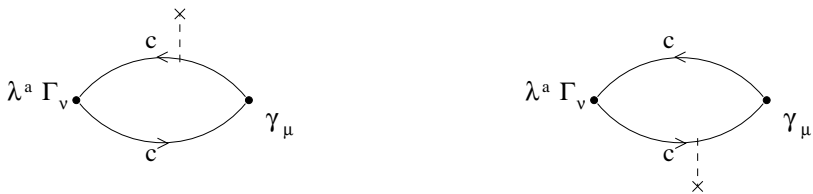,width=\textwidth,bbllx=0pt,bblly=0pt,bburx=470pt,%
bbury=100pt,clip=}
}
\caption{\it Diagrams determining the Wilson coefficients
in the OPE (\ref{expan}).}
\end{figure}

The higher-dimensional terms of the expansion 
denoted by the ellipses in (\ref{expan}) involve operators with derivatives 
of the gluon field and/or 
multiple gluon-field operators. As discussed in ref. \cite{KRSW}, 
for the physical $b$ and $c$ masses, these terms are 
expected to contribute at the level of $ 10\%$ or less. 
We neglect them  here. 

Substitution of (\ref{expan}) in (\ref{corr}) yields 
\be
A(p^2)= -g_{BK}(p^2)I_c(p^2)~,
\label{matr}
\ee
where the effective form factor $g_{BK}$ is defined by 
the matrix element 
\be
t_{\mu\nu\alpha\beta}
\langle K(q) \mid 
\bar{s} \Gamma^\nu
g_s\widetilde{G}^{\alpha\beta}b
\mid B(p+q)\rangle=\left( p^2 q_\mu -(p\cdot q)p_\mu \right)
g_{BK}(p^2)
\label{ope2}
\ee 
with $\widetilde{G}^{\alpha\beta} = \frac{\lambda^a}2\widetilde{G}^{a\alpha\beta}$.
Note that the relevant scale for $A$  is the scale 
of OPE, that is $\mu \simeq 2m_c$. 
The calculation of  the form factor $g_{BK}(p^2)$ requires some 
nonperturbative method. In the next section 
we will estimate it using light-cone
sum rules. 


\section{Light-cone sum rule for the form factor $g_{BK}$}

To evaluate the matrix element (\ref{ope2})
we proceed as in the calculation of the $B\to \pi,K$ 
form factors in ref. \cite{BKR} (see also the reviews ref. \cite{KR,Braun}). 
Considering the correlation function 
$$
F_{\mu}(p,q)=
it_{\mu\nu\alpha\beta}  
\int d^4x e^{ipx}\langle K(q)\mid T\{
\bar{s}(x) \Gamma^\nu g_s\widetilde{G}^{\alpha\beta}(x)b(x),
\bar{b}(0)i\gamma_5 d(0)\}\mid 0\rangle
$$
\be
= \left( p^2 q_\mu -(p\cdot q)p_\mu \right)F(p^2, (p+q)^2)~,
\label{corr2}
\ee
we insert a complete set
of intermediate states with $B$-meson quantum numbers
between the currents. This yields:
$$
F_{\mu} (p,q)=
t_{\mu\nu\alpha\beta} \Bigg(\frac{\langle K \mid \bar{s}
\Gamma^\nu g_s\widetilde{G}^{\alpha\beta}b \mid B\rangle
\langle B\mid \bar{b} i\gamma_5 d \mid 0\rangle}{m_B^2-(p+q)^2}
$$
\be
\label{2a}
+\sum_{h_B}\frac{\langle K \mid \bar{s}\Gamma^\nu
g_s\widetilde{G}^{\alpha\beta}b \mid h_B\rangle
\langle h_B \mid \bar{b}i\gamma_5d\mid 0\rangle}{m_{h_B}^2-(p+q)^2}
\Bigg)~,
\label{dispa}
\ee
where the ground state contribution 
contains the matrix element of interest. Using  
(\ref{ope2}) and 
\be
m_b\langle 0 \mid\bar{b}i\gamma_5 d\mid B\rangle =m_B^2f_B~,
\label{fB2}
\ee
where $f_B$ is the $B$ meson decay constant, and 
representing the sum over excited and 
continuum states by a dispersion
integral, one finds 
\be
F(p^2, (p+q)^2)= \frac{m_B^2f_Bg_{BK}(p^2)}{m_b(m_B^2-(p+q)^2)}
+ \int\limits_{s_0^h}^{\infty} \frac{\rho^{h_B}(p^2,s)ds}{s-(p+q)^2}~.
\label{disp1}
\ee
In the momentum region $(p+q)^2 \ll m_b^2$  and $p^2 \ll 4m_c^2$, 
the amplitude $F$ 
can also be calculated with the help of OPE. 
The result can be brought in the  form of a dispersion 
integral:
\be
F_{QCD}(p^2,(p+q)^2)=\frac1{\pi}\int\limits_{m_b^2}^{\infty}
ds ~\frac{ \mbox{Im} F_{QCD}(p^2, s)}{s-(p+q)^2}.
\label{qcd}
\ee
Furthermore, the spectral density of the higher states 
in (\ref{disp1}) is approximated by 
\be
\rho^{h_B}(p^2,s) \Theta(s - s_0^h)
 = \frac1\pi \mbox{Im} F_{QCD}(p^2, s) \Theta( s-s_0^B) ~,
\label{rhoh5}
\ee
as suggested by quark-hadron duality.
With (\ref{rhoh5}) it is then straightforward to subtract
the contribution of the excited and continuum states 
in the equation given by (\ref{disp1}) and (\ref{qcd})
and solve this equation for $g_{BK}$.
After performing the Borel transformation in $(p+q)^2$, 
the solution reads:
\begin{eqnarray}
g_{BK}(p^2)=\frac{m_b}{\pi f_B m^2_B}\int\limits_{m_b^2}^{s_0^B} 
\mbox{Im} F_{QCD}(p^2, s)
\exp\left(\frac{m^2_B-s}{M^2}\right )ds~.
\label{sr}
\end{eqnarray}

In contrast to conventional sum rules based on the short-distance 
OPE in terms of local operators, here an
expansion in terms of nonlocal operators near the light-cone, $x^2\simeq 0$,
is used to calculate the invariant amplitude $F_{QCD}$.
Contraction of the $b$-quark fields in the correlation function 
(\ref{corr2}) 
and use of the free $b$-quark propagator 
leads to the diagram Fig. 2a. In this approximation, $F_{QCD}$ 
is given by a set of bilocal matrix 
elements $ \langle K |\bar{s}(x)\Gamma_a g_s G_{\alpha\beta}(x)d(0)|0\rangle$
where $\Gamma_a$ denotes a combination of Dirac matrices.
These matrix elements are parametrized in terms of three-particle 
light-cone wave functions of different twist: 
\bq
\langle K |\bar{s}(x)
\sigma_{\mu\nu}\gamma_5g_s G_{\alpha\beta}(x)d(0)|0\rangle 
= if_{3K}[(q_\alpha q_\mu g_{\beta\nu}-q_\beta q_\mu g_{\alpha\nu})
\nonumber\\
-(q_\alpha q_\nu g_{\beta\mu}-q_\beta q_\nu g_{\alpha\mu})]
\int{\cal D}\alpha_i\,\varphi_{3K}(\alpha_i)e^{iqx(\alpha_1+\alpha_3)}\,,
\label{29}
\eq
\bq
\langle K |\bar{s}(x)
\gamma_\mu\gamma_5 g_sG_{\alpha\beta}(x)d(0)|0\rangle
=f_K\Bigg[ q_\beta\left( g_{\alpha\mu}
-\frac{x_\alpha q_\mu}{qx}\right) 
\nonumber
\\
-q_\alpha\left(g_{\beta\mu}-\frac{x_\beta q_\mu}{qx}\right)\Bigg]
\int{\cal D}\alpha_i\varphi_{\perp K} (\alpha_i)e^{iqx(\alpha_1+\alpha_3)}
\nonumber
\\
+f_K\frac{q_\mu}{qx}(q_\alpha x_\beta -q_\beta x_\alpha )
\int{\cal D}\alpha_i\,\varphi_{\parallel K} (\alpha_i)e^{iqx
(\alpha_1+\alpha_3)}\,,
\label{30}
\eq

\bq
\langle K |\bar{s}(x)\gamma_\mu g_s\widetilde{G}_{\alpha\beta}(x)d(0)|0\rangle
=if_K\Bigg[ q_\beta\left( g_{\alpha\mu}-\frac{x_\alpha q_\mu}{qx}\right) 
\nonumber
\\
-q_\alpha\left( g_{\beta\mu}-\frac{x_\beta q_\mu}{qx}\right)\Bigg]
\int{\cal D}\alpha_i\,\widetilde{\varphi}_{\perp K}(\alpha_i)e^{iqx
(\alpha_1+\alpha_3)}
\hspace{1.4cm}{}
\nonumber
\\
+if_K\frac{q_\mu}{qx}(q_\alpha x_\beta -q_\beta x_\alpha )
\int{\cal D}\alpha_i\,\widetilde{\varphi}_{\parallel K} (\alpha_i)
e^{iqx(\alpha_1+\alpha_3)}
\label{31}
\eq
with 
${\cal D}\alpha_i= d\alpha_1 d\alpha_2 d\alpha_3 \delta(1-\alpha_1-\alpha_2
-\alpha_3)$. 
The wave function 
$\varphi_{3K}(\alpha_i)=\varphi_{3K}(\alpha_1,\alpha_2,\alpha_3)$ 
has twist 3, while  
$\varphi_{\perp K}$, $\varphi_{\parallel K}$,
$\widetilde{\varphi}_{\perp K}$ and $\widetilde{\varphi}_{\parallel K}$ 
are all of twist 4.
In the following, we consider the $SU(3)$-symmetry limit where 
$f_{3K}=f_{3\pi}$, $\varphi_{3K}=\varphi_{3\pi}$,
$\varphi_{\perp K}= \varphi_{\perp \pi}$, etc. 
The explicit expressions for the pion wave functions can be found in 
ref. \cite{KR}.
Higher Fock-space components of the $K$ meson wave function are 
neglected as well as perturbative corrections of the kind 
shown in Fig. 2b. Omitting further details of the calculation 
we present the final result: 
$$
F(p^2,(p+q)^2)= 2f_{3K}(qp)
\int {\cal D}\alpha_i 
\frac{\varphi_{3K}(\alpha_i)}{m_b^2-(p+(\alpha_1+\alpha_3)q)^2}
$$ 
$$
+2f_{K}m_b\int\limits^1_0 \frac{du}{m_b^2-(p+uq)^2} 
\Bigg\{\int\limits^u_0 
d\alpha_3 \left( 3\widetilde{\varphi}_{\perp K}(\alpha_i)
-\widetilde{\varphi}_{\parallel K}(\alpha_i)\right)_
{\stackrel{{\scriptstyle \alpha_1=u-\alpha_3}}{\alpha_2=1-u}}
$$
\bq
+
\frac1u\left( \frac{m_b^2-p^2}{m_b^2-(p+uq)^2}-1 \right)
\int\limits^u_0dv\int\limits^v_0 \hspace{-2pt}
d\alpha_3 \left(\widetilde{\varphi}_{\perp K}(\alpha_i)+
\widetilde{\varphi}_{\parallel K}(\alpha_i)\right)_
{\stackrel{{\scriptstyle \alpha_1=v-\alpha_3}}{\alpha_2=1-v}}
\hspace{-5pt}\Bigg\}.\hspace{2pt}
\label{tw4}
\eq
Substituting the above result in (\ref{sr})
yields the light-cone sum rule:
$$
g_{BK}(p^2)= \frac{m_b^2}{f_Bm_B^2}\int\limits_{\Delta}^1
\frac{du}u
\exp\left ( \frac{m_B^2}{M^2}-\frac{m_b^2-p^2(1-u)}{uM^2} \right)
$$
$$
\times\Bigg\{
\int\limits^u_0 d\alpha_3 \left( f_{3K}\frac{m_b^2-p^2}{m_bu}
\varphi_{3K}(\alpha_i)+f_\pi \left(3\widetilde{\varphi}_{\perp K}(\alpha_i)
-\widetilde{\varphi}_{\parallel K }(\alpha_i)\right)
\right)_
{\stackrel{{\scriptstyle \alpha_1=u-\alpha_3}}{\alpha_2=1-u}}
$$
\bq
+\frac1u\left( \frac{m_b^2-p^2}{uM^2}-1 \right)
\int\limits^u_0dv~\int\limits^v_0
d\alpha_3 \left(\widetilde{\varphi}_{\perp K}(\alpha_i)
+\widetilde{\varphi}_{\parallel K}(\alpha_i)\right)_
{\stackrel{{\scriptstyle \alpha_1=v-\alpha_3}}{\alpha_2=1-v}}
\Bigg\}~, 
\label{srfinal}
\eq
where $\Delta=m_b^2-p^2/(s_0^B-p^2)$.
In the above estimate the scale of the form factor 
$g_{BK}$ is set by the Borel mass $M\simeq \sqrt{m_B^2-m_b^2}$. 

\begin{figure}[ht]
\mbox{
\epsfig{file=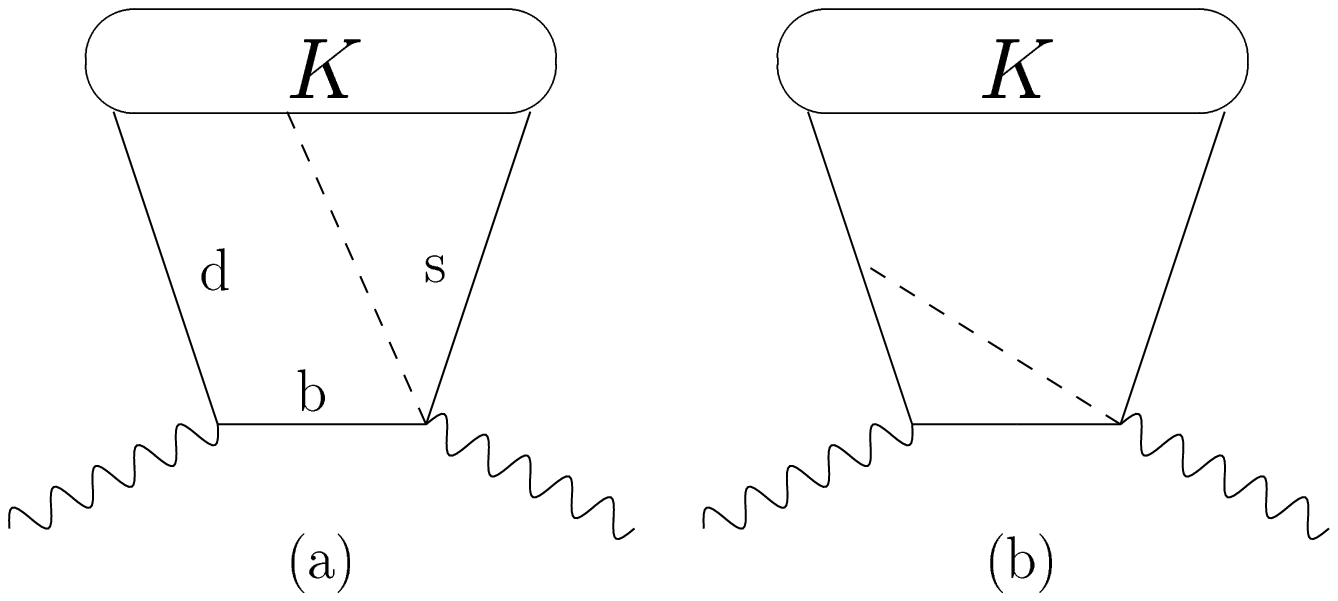,width=\textwidth,bbllx=0pt,bblly=260pt,bburx=600pt,%
bbury=570pt,clip=}
}
\caption{\it Diagrammatic representation of the correlation funciton 
(\ref{corr2}). Solid lines represent quarks, dashed line gluons, wavy 
lines external currents, and ovals 
light-cone wave functions of the kaon.}
\end{figure}

\section{Estimate of the nonfactorizable amplitude in $B\to J/\psi K$}

From (\ref{matr}) and (\ref{srfinal}) 
we get an estimate for the invariant amplitude $A(p^2)$
which is related to $\tilde{f}_{B\psi K}$ by (\ref{disppsi}).
In order to suppress the contribution of excited charmonium
and nonresonant states in the latter dispersion 
relation and to eliminate subtraction terms we take the $n$-th 
derivative of (\ref{disppsi}) at $p^2=0$:
$$
\tilde{f}_{B\psi K}+ \left(\frac{f_{\psi'}}{f_{\psi}}\right)^2
\left(\frac{m_\psi}{m_{\psi'}} \right)^{2n+2}\tilde{f}_{B\psi' K}
$$
\be
+ \frac{m_\psi^{2n+2}}{2f_\psi^2} \int\limits_{4m_D^2}^{\infty}
\frac{\rho^h(s)~ds}{s^{n+1}}
=- 
\frac{m_\psi^{2n+2}}{2n! f_\psi^2}\frac{d^n}{dp^{2n}}\Big[g_{BK}(p^2)I_c(p^2)
\Big]_{p^2=0}~.
\label{disppsi1}
\ee
Since the influence of higher-dimensional operators 
neglected in (\ref{expan}) increases with $n$ we restrict 
the numerical analysis with  $n \leq 8$.
For the charmed quark mass we take
$m_c=1.3 $ GeV and for decay constants the values 
$f_\psi=405 $ MeV, $f_{\psi'}/f_{\psi} \simeq 0.7 $
extracted from the measured 
leptonic widths of the $J/\psi$ and $\psi'$  
\cite{PDG}. 
The values of parameters appearing in the light-cone sum rule 
for $g_{BK}$ are given in ref. \cite{KR}.

The contribution from the $\psi'$ to the sum rule 
(\ref{disppsi1})  is suppressed relative to the $J/\psi$ 
contribution roughly by a factor 10 at $n=3$ and a factor $50$ 
at $n=8$. It is therefore reasonable to neglect the integral over
still heavier states in (\ref{disppsi1}) altogether. 
From the remaining sum rule one can get a first rough estimate of 
$\tilde{f}_{B\psi K}$ by taking $n \geq 6$ and dropping the $\psi'$ term.
This yields 
\be
\tilde{f}_{B\psi K}(\mu) = -(0.04, ~0.05, ~0.07)~~\mbox{at}~~n=6,7,8~,
\label{number}
\ee
where $\mu\simeq 2m_c = 2.6$ GeV  which accidentally coincides 
with the scale $M \simeq \sqrt{m_B^2-m_b^2}= 2.4$ GeV
of the form factor $g_{BK}$ used to get (\ref{number}). 
For a more elaborated estimate one may 
fit the amplitudes $\tilde{f}_{B\psi K}$ and $ \tilde{f}_{B\psi' K}$ 
to the   moments (\ref{disppsi1}) for  $n=2\div 8$.
This gives $\tilde{f}_{B\psi K}(\mu) = -0.06$  
and $ \tilde{f}_{B\psi' K}(\mu) = +0.3$.

\section{Discussion}

QCD sum rule techniques together with light-cone wave 
functions provide new ways to 
go beyond factorization in exclusive nonleptonic 
decays of heavy mesons. Within this approach, we 
have estimated the nonfactorizable amplitude 
(\ref{def}) for the decay $B\to J/\psi K$ . The 
result (\ref{number}) agrees with an
earlier estimate (\ref{number4pt}) 
obtained from a four-point sum rule \cite{KR}.
Furthermore, it is interesting to note that 
the nonfactorizable amplitude  $\tilde{f}_{B\psi' K}$ for $B\to \psi'K$, 
differs from   $ \tilde{f}_{B\psi K} $
both in sign and magnitude. This finding underlines the 
general expectation of nonuniversality 
of the  coefficients $a_2$ in the sum rule approach.

Although it is encouraging that both methods give similar results,
there are still unknown uncertainties.
They arise from several sources: neglect of 
higher-dimensional operators, unknown  perturbative corrections,
a crude model for the hadronic spectral density 
of higher states in the $J/\psi$ channel, and numerical uncertainties
on the parameters and wave functions.
The uncertainty on $\tilde{f}_{B\psi' K}$ is 
obviously larger than the one on  $\tilde{f}_{B\psi K}$,
because the former has a much smaller coefficient in (\ref{disppsi1})
than the latter.

Our final comment concerns the original applications in refs. 
\cite{BS93,Halperin} 
of this method to $B\to D\pi$ .
Let us consider the  $\bar{B}^0 \to D^0\pi^0$ mode. 
The nonfactorizable amplitude 
\be
\langle
D^0\pi^0 \mid \widetilde{O}_2'\mid \bar{B}^o \rangle
\label{operD}
\ee
with 
\be
\widetilde{O}_2'=(\bar{c}\Gamma^\rho \frac{\lambda^a}{2}u)
(\bar{d}\Gamma_\rho \frac{\lambda^a}{2}b)
\label{operD1}
\ee
has been estimated in ref. \cite{Halperin} 
from the correlation function 
$\langle \pi^0| T\{\bar{u}\gamma_\mu\gamma_5 c, 
\widetilde{O}_2'\}|\bar{B}^0\rangle$, $\bar{u}\gamma_\mu\gamma_5 c$
being the 
generating current of $D^0$.
Contraction of all quark fields in the operator product 
\be
T\{\bar{u}(x)\gamma_\mu\gamma_5 c(x),\bar{c}(0)
\Gamma^\rho \frac{\lambda^a}{2}u(0)\}
\label{opeDD}
\ee 
leads to the gluon-field operator, as shown in section 2.  
However, the OPE of (\ref{opeDD}) also contains 
the operator $\bar{u}\frac{\lambda^a}2\Gamma_r u $ 
resulting from contraction of the heavy-quark fields alone. 
This operator contributes to the sum rule 
for  (\ref{operD}) through matrix elements of effective 
four-quark operators 
$\langle \pi^0 \mid (\bar{u}\frac{\lambda^a}2\Gamma_r u) 
(\bar{d}\Gamma_\rho  \frac{\lambda^a}2 b) \mid B \rangle$.
These contributions which have not been taken into
account in refs. \cite{BS93,Halperin} may significantly influence 
the results for the nonfactorizable amplitudes in $B\to D\pi$.   

\section*{Acknowledgments}

We are grateful to C.W. Winhart for collaboration 
on the subject of this talk.
A.K. wants to thank M. Shifman, A. Vainshtein and M. Voloshin
for their kind invitation to a fruitful Workshop. 
This work was supported by the Bundesministerium f\"ur Bildung,
Wissenschaft, Forschung und Technologie, Bonn, Germany,
Contract 05 7WZ91P (0).

\end{document}